\documentclass[12pt]{article}
\usepackage{epsfig}
\usepackage{cite}

\newcommand{\mysection}{\setcounter{equation}{0}\section}

\def\beq{\begin{equation}}
\def\eeq{\end{equation}}
\def\beqa{\begin{eqnarray}}
\def\eeqa{\end{eqnarray}}
 
\newlength{\dinwidth} \newlength{\dinmargin}
\begin{document}

\begin{center}
{\Large \bf Higher-order radiative corrections for $b{\bar b} \rightarrow H^- W^+$}
\end{center}
\vspace{2mm}
\begin{center}
{\large Nikolaos Kidonakis}\\
\vspace{2mm}
{\it Department of Physics, Kennesaw State University,\\
Kennesaw, GA 30144, USA}
\end{center}
 
\begin{abstract}
I present higher-order radiative corrections from collinear and soft gluon 
emission for the associated production of a charged Higgs boson with a $W$ 
boson. The calculation uses expressions from resummation at 
next-to-leading-logarithm accuracy.  
From the resummed cross section I derive analytical formulas at 
approximate NNLO and N$^3$LO.
Total cross sections are presented for  
the process $b{\bar b} \rightarrow H^- W^+$ at various LHC energies. 
The transverse-momentum and rapidity distributions of the charged Higgs 
boson are also calculated.
\end{abstract}
 
\mysection{Introduction}
 
Higgs bosons play a central role in both the Standard Model and in searches for new physics. Two-Higgs-doublet models in new physics scenarios, such as the Minimal Supersymmetric Standard Model, involve charged Higgs bosons in addition to neutral ones. One of the Higgs doublets gives mass to up-type fermions while the other to down-type fermions, with the ratio of the vacuum expectation values for the two doublets denoted by $\tan\beta$. Two charged Higgs bosons, 
$H^+$ and $H^-$, appear in such models.

An important charged Higgs production process at LHC energies is the associated production of a charged Higgs boson with a $W$ boson, which may proceed via the partonic process $b{\bar b} \rightarrow H^- W^+$ or $b{\bar b} \rightarrow H^+ W^-$. This process was studied in Refs. \cite{DHKR,DK,YLJZ,ZMJHW,BHK,HZ,ABK,ZLL,EHR,GLL,MH,BTW,NHN,AGKMSY,ARD,BGLLS,EPS,LWY}, and various kinds of radiative corrections were calculated in those works. There is good potential for the LHC to discover charged Higgs bosons via this process, so it is useful to calculate higher-order corrections that may enhance the cross section.

An important set of higher-order corrections is due to soft-gluon emission, dominant near partonic threshold; another is due to collinear gluon emission. These corrections can in principle be resummed, and the resummation formalism can be used to construct approximate higher-order results.

In this paper I present a first study of collinear and soft-gluon resummation for the associated production of a charged Higgs boson with a $W$ boson via $b$-quark annihilation. Since the charged Higgs is presumably very massive, its possible production at the LHC would be a near-threshold process.

I employ the resummation formalism that has been used for several related processes, including charged Higgs production in association with a top quark \cite{NKcH,NKtWH}, neutral Higgs production via $b{\bar b}$ annihilation \cite{NKbbH}, $W$ or $Z$ production at large transverse momentum \cite{NKWZ}, top-quark production in association with a $W$ boson \cite{NKtWH,NKtW,NKtop}, and top-antitop pair production \cite{NKtop,NKttbar}.

In the next section we discuss collinear and soft-gluon corrections and present their resummation. Using the expansion of the resummed cross section at next-to-leading order (NLO), next-to-next-to-leading order (NNLO), and next-to-next-to-next-to-leading order (N$^3$LO), we derive approximate NLO (aNLO), approximate NNLO (aNNLO), and approximate N$^3$LO (aN$^3$LO), cross sections. In Section 3 we present results for $H^- W^+$ total cross sections at LHC energies. In Section 4 we present results for the charged Higgs transverse momentum and rapidity distribution in this process. We conclude in Section 5.

\mysection{Collinear and soft-gluon resummation for $b{\bar b} \rightarrow H^- W^+$}

For the process $b{\bar b} \rightarrow H^- W^+$, involving bottom quarks in 
the initial state, we assign the momenta
\beq
b(p_1)\, + \, {\bar b}\, (p_2) \rightarrow H^-(p_3)\, + W^+(p_4) \, , 
\eeq
and define the kinematical variables 
$s=(p_1+p_2)^2$, $t=(p_1-p_3)^2$, $t_1=t-m_H^2$, $t_2=t-m_W^2$, 
$u=(p_2-p_3)^2$, $u_1=u-m_H^2$, and $u_2=u-m_W^2$, where $m_H$ is the 
charged Higgs mass and  $m_W$ is the $W$-boson mass while the $b$-quark 
mass is taken to be 0. We note that we work in the five-flavor scheme where the
$b$-quark is treated as a parton in the proton. 

We also define the variable $s_4=s+t_1+u_2$, which 
measures distance from partonic threshold where there is no energy for 
additional emission; however, even when $s_4=0$ the charged Higgs boson and 
the $W$ boson are not constrained to be produced at rest. 
We note that identical considerations apply to $H^+ W^-$ production.

Radiative corrections, including collinear and soft-gluon corrections, 
appear at each order in the perturbative expansion of the cross section. 
The resummation of these corrections in our formalism is performed for the double-differential cross section in single-particle-inclusive (1PI) kinematics, in terms of the variable $s_4$.
We note that while resummation for colorless final states is well established, previous studies have not been done in 1PI kinematics but have instead used the more inclusive variable $z=M^2/s$, where $M$ is the invariant mass of the final state. Therefore, the present work is distinct from other work on Higgs or other electroweak final states. Using the $s_4$ resummation introduces several additional new terms in the expressions for the higher-order corrections, as we will discuss later. Furthermore, our 1PI resummation formalism allows the calculation of higher-order soft-gluon contributions to the Higgs transverse-momentum and rapidity distributions, something which is not possible with the resummation in invariant mass.

The soft-gluon terms are plus distributions of logarithms of $s_4$, 
$[\ln^k(s_4/m_H^2)/s_4]_+$, with $k$ an integer ranging from 0 
to $2n-1$ for the $n$th order corrections in the strong coupling, $\alpha_s$.
The plus distributions are defined by their integrals with functions $f$, 
which in our case involve perturbative coefficients and parton distribution 
functions (pdf) as discussed later, via the expression 
\beqa
\int_0^{s_4^{max}} ds_4 \, \left[\frac{\ln^k(s_4/m_H^2)}
{s_4}\right]_{+} f(s_4) &=&
\int_0^{s_4^{max}} ds_4 \frac{\ln^k(s_4/m_H^2)}{s_4} [f(s_4) - f(0)]
\nonumber \\ &&
{}+\frac{1}{k+1} \ln^{k+1}\left(\frac{s_4^{max}}{m_H^2}\right) f(0) \, .
\label{plus}
\eeqa

In addition, further logarithmic terms of the form $(1/m_H^2) \ln^k(s_4/m_H^2)$, of collinear origin, also appear in the perturbative expansion. These collinear terms are fully known only at leading logarithmic accuracy. In this paper we provide the first analytical and numerical study of such terms in 1PI kinematics with the $s_4$ variable.

Resummation of collinear and soft-gluon contributions follows from  the factorization of the cross section into various functions that describe collinear and soft emission in the partonic process. Taking moments of the partonic scattering cross section, ${\hat \sigma}(N)=\int (ds_4/s) \;  e^{-N s_4/s} {\hat \sigma}(s_4)$, with $N$ the moment variable, we write a factorized expression in $4-\epsilon$ dimensions:
\beq
{\hat \sigma}^{H^- W^+}(N,\epsilon)= 
\left( \prod_{i=b,{\bar b}} J_i\left (N,\mu,\epsilon \right) \right)
H^{H^- W^+} \left(\alpha_s(\mu)\right)\; S^{H^- W^+} 
\left(\frac{m_H}{N \mu},\alpha_s(\mu) \right)\;
\label{factsigma}
\eeq 
where $\mu$ is the scale, $J_i$ are jet functions that describe 
soft and collinear emission from the incoming $b$ and ${\bar b}$ quarks, 
$H^{H^- W^+}$ is the hard-scattering function, and 
$S^{H^- W^+}$ is the soft-gluon function for non-collinear soft-gluon emission.
The lowest-order cross section is given by the product of the lowest-order 
hard and soft functions.

The soft function $S^{H^- W^+}$ requires renormalization, and its $N$-dependence 
can be resummed via renormalization group evolution. 
Thus, $S^{H^- W^+}$ satisfies the renormalization group equation
\beq
\left(\mu \frac{\partial}{\partial \mu}
+\beta(g_s, \epsilon)\frac{\partial}{\partial g_s}\right)\,S^{H^- W^+}
=-2 \, S^{H^- W^+} \, \Gamma_S^{H^- W^+}
\eeq
where $g_s^2=4\pi\alpha_s$; 
$\beta(g_s, \epsilon)=-g_s \epsilon/2 + \beta(g_s)$ 
with $\beta(g_s)$ the QCD beta function; and 
$\Gamma_S^{H^- W^+}$ is the soft anomalous dimension
that controls the evolution of the soft-gluon function $S^{H^- W^+}$.

The evolution of the soft and jet functions provides resummed expressions 
for the cross section \cite{NKcH,NKtWH,NKbbH,NKWZ,NKtW,NKtop,NKttbar}.
For $H^- W^+$  production the resummed partonic cross section in moment space 
is given by  
\beqa
{\hat{\sigma}}_{\rm res}^{H^- W^+}(N) &=&   
\exp\left[\sum_{i=b,{\bar b}} E_i(N_i)\right]
H^{H^- W^+} \left(\alpha_s(\sqrt{s})\right) \;
S^{H^- W^+}\left(\alpha_s(\sqrt{s}/{\tilde N'})
\right) 
\nonumber \\ &&
\times \exp \left[2\int_{\sqrt{s}}^{{\sqrt{s}}/{\tilde N'}} 
\frac{d\mu}{\mu}\; \Gamma_S^{H^- W^+}
\left(\alpha_s(\mu)\right)\right]  \, .
\label{resHS}
\eeqa
The first exponent \cite{GS87,CT89} in Eq. (\ref{resHS}) resums soft and 
collinear corrections from the incoming $b$ and ${\bar b}$ quarks and is 
well known (see \cite{NKtWH,NKbbH,NKtop} for details). Since the resummation
is performed in 1PI kinematics, we have $N_b=N(-u_2/m_H^2)$ and 
$N_{\bar b}=N(-t_2/m_H^2)$, and this generates 
logarithms involving $t_2$ and $u_2$ in the fixed-order expansions. This is an 
important point, as no such terms appear in invariant-mass resummations, 
for which $N_b=N_{\bar b}=N$.  

The specific forms of the expressions for the individual terms 
in Eq. (\ref{resHS}) depend on the gauge, although the overall result 
for the resummed cross section of course does not.
In Feynman gauge the one-loop soft anomalous dimension for 
$b{\bar b} \rightarrow H^- W^+$ vanishes; in axial gauge it is 
$(\alpha_s/\pi) C_F$, where $C_F=(N_c^2-1)/(2N_c)$ with $N_c=3$ the number of 
colors. 
We calculate the soft-gluon corrections at next-to-leading-logarithm 
accuracy. However, as mentioned previously, only the leading collinear 
corrections are fully known.

We expand the resummed cross section, Eq. (\ref{resHS}), in $\alpha_s$, 
and then we invert to momentum space. We provide explicit analytical results 
through third order for the collinear and soft-gluon corrections. 

The NLO collinear and soft-gluon corrections from the resummation are
\beqa
\frac{d^2{\hat{\sigma}}^{(1)}}{dt \, du} &=& \frac{\pi \alpha^2 \, m_t^4 \, 
\cot^2\beta}{48 \sin^4\theta_W \, m_W^4 \, s^2 \, t_1^2} 
\left(m_W^2 s +t_2 \, u_2\right) \frac{\alpha_s(\mu_R)}{\pi} C_F
\left\{-\frac{4}{m_H^2} \ln\left(\frac{s_4}{m_H^2}\right) \right. 
\nonumber \\ && \hspace{-15mm} 
{}+4 \left[\frac{\ln(s_4/m_H^2)}{s_4}\right]_+
-2 \left[\ln\left(\frac{t_2 \, u_2}{m_H^4}\right)
+\ln\left(\frac{\mu_F^2}{s}\right)\right] \left[\frac{1}{s_4}\right]_+
\nonumber \\ && \hspace{-15mm} \left.
+\left[\ln\left(\frac{t_2 \, u_2}{m_H^4}\right)-\frac{3}{2}\right]
\ln\left(\frac{\mu_F^2}{m_H^2}\right)\,  \delta(s_4)
\right\}
\label{NLO}
\eeqa
where $\alpha=e^2/(4\pi)$, $\theta_W$ is the weak mixing angle,  
$\mu_R$ is the renormalization scale, and $\mu_F$ is the factorization scale.
We note that the logarithmic terms involving the variables $t_2$ and $u_2$ 
in the above expression arise from the 1PI nature of our resummation and 
would not appear in an invariant-mass resummation.

The NNLO collinear and soft-gluon corrections from the resummation are
\beqa
\frac{d^2{\hat{\sigma}}^{(2)}}{dt \, du}&=&
\frac{\pi \alpha^2 \, m_t^4 \, \cot^2\beta}
{48 \sin^4\theta_W \, m_W^4 \, s^2 \, t_1^2} 
\left(m_W^2 s +t_2 \, u_2\right) 
\frac{\alpha_s^2(\mu_R)}{\pi^2} C_F
\nonumber \\ && \hspace{-18mm} \times 
\left\{-8 C_F\ \frac{1}{m_H^2} \ln^3\left(\frac{s_4}{m_H^2}\right) 
+8 C_F\, \left[\frac{\ln^3(s_4/m_H^2)}{s_4}\right]_+ \right.
\nonumber \\ && \hspace{-18mm} 
{}+\left[-12 C_F \left(\ln\left(\frac{t_2 \, u_2}{m_H^4}\right)
+\ln\left(\frac{\mu_F^2}{s}\right)\right)  
-\frac{11}{3} C_A +\frac{2}{3} n_f\right]  
\left[\frac{\ln^2(s_4/m_H^2)}{s_4}\right]_+ 
\nonumber \\ && \hspace{-18mm} 
{}+\left[4C_F \ln^2\left(\frac{\mu_F^2}{m_H^2}\right)
+C_F \left(12 \ln\left(\frac{t_2 \, u_2}{m_H^4}\right)
+8\ln\left(\frac{m_H^2}{s}\right)-6\right)
\ln\left(\frac{\mu_F^2}{m_H^2}\right)\right.
\nonumber \\ && \hspace{-13mm} \left.
{}+\left(\frac{11}{3} C_A -\frac{2}{3} n_f\right)
\ln\left(\frac{\mu_R^2}{m_H^2}\right)\right]
\left[\frac{\ln(s_4/m_H^2)}{s_4}\right]_+ 
\nonumber \\ && \hspace{-18mm} 
{}+\left[\left(-2 C_F\ln\left(\frac{t_2 \, u_2}{m_H^4}\right)
+3 C_F+\frac{11}{12} C_A -\frac{n_f}{6} \right) \right.
\ln^2\left(\frac{\mu_F^2}{m_H^2}\right)
\nonumber \\ && \hspace{-13mm} \left. \left.
{}-\left(\frac{11}{6} C_A -\frac{n_f}{3} \right)
\ln\left(\frac{\mu_F^2}{m_H^2}\right) \ln\left(\frac{\mu_R^2}{m_H^2}\right) 
\right]\left[\frac{1}{s_4}\right]_+ \right\}
\label{NNLO}
\eeqa
where $C_A=N_c$, and $n_f=5$ is the number of light-quark flavors.
Again, the logarithmic terms involving the variables $t_2$ and $u_2$ 
in the above expression arise from the 1PI nature of the resummation.

Equation (\ref{NNLO}) can be written more compactly as 
\beqa
\frac{d^2{\hat{\sigma}}^{(2)}}{dt \, du}&=& F_{LO} \frac{\alpha_s^2}{\pi^2}
\left\{-C_3^{(2)} \frac{1}{m_H^2} \ln^3\left(\frac{s_4}{m_H^2}\right) 
+\sum_{k=0}^3 C_k^{(2)} \left[\frac{\ln^k(s_4/m_H^2)}{s_4}\right]_+ \right\}
\label{NNLOcompact}
\eeqa
where $F_{LO}$ denotes the overall leading-order factor and the $C_k^{(2)}$ are 
coefficients of the logarithms, and they can be read off by comparing 
Eq. (\ref{NNLOcompact}) with Eq. (\ref{NNLO}), e.g. $C_3^{(2)}=8 C_F^2$. 
This compact form for the aNNLO corrections will be useful in the next section.

Finally, one can consider the contribution of even higher-order corrections although not all logarithms can be determined. The N$^3$LO collinear and soft-gluon corrections from the resummation are
\beqa
\frac{d^2{\hat{\sigma}}^{(3)}}{dt \, du}&=& F_{LO} \frac{\alpha_s^3}{\pi^3}
\left\{-C_5^{(3)} \frac{1}{m_H^2} \ln^5\left(\frac{s_4}{m_H^2}\right) 
+\sum_{k=0}^5 C_k^{(3)} \left[\frac{\ln^k(s_4/m_H^2)}{s_4}\right]_+ \right\}
\label{N3LOcompact}
\eeqa
where the $C_k^{(3)}$ are coefficients of the logarithms. We have 
$C_5^{(3)}=8 C_F^3$, 
\beq
C_4^{(3)}=-20 C_F^3\left[\ln\left(\frac{t_2 \, u_2}{m_H^4}\right)
+\ln\left(\frac{\mu_F^2}{s}\right)\right] -\frac{10}{3}\beta_0 C_F^2 \, ,
\eeq 
\beqa
C_3^{(3)}&=&-64 C_F^3 \zeta_2 +8 C_F^3 \ln\left(\frac{\mu_F^2}{m_H^2}\right)
\left[2\ln\left(\frac{\mu_F^2}{m_H^2}\right)+4\ln\left(\frac{m_H^2}{s}\right)
+5 \ln\left(\frac{t_2 \, u_2}{m_H^4}\right)-\frac{3}{2}\right]
\nonumber \\ &&
{}+4C_F^2 \beta_0 \left[\frac{2}{3}\ln\left(\frac{\mu_F^2}{m_H^2}\right)
+\ln\left(\frac{\mu_R^2}{m_H^2}\right)\right]
\eeqa
\beqa
C_2^{(3)}&=&160 C_F^3 \zeta_3-4C_F^3 \ln^3\left(\frac{\mu_F^2}{m_H^2}\right)
-12 C_F^3 \ln^2\left(\frac{\mu_F^2}{m_H^2}\right)
\left[2\ln\left(\frac{t_2 \, u_2}{m_H^4}\right)+\ln\left(\frac{m_H^2}{s}\right)
-\frac{3}{2}\right]
\nonumber \\ &&
{}+96C_F^3 \zeta_2 \left[\ln\left(\frac{t_2 \, u_2}{m_H^4}\right)
+\ln\left(\frac{\mu_F^2}{s}\right)\right]
-6\beta_0 C_F^2 \ln\left(\frac{\mu_F^2}{m_H^2}\right)
\ln\left(\frac{\mu_R^2}{m_H^2}\right)
+\frac{3}{2}\beta_0 C_F^2 \ln^2\left(\frac{\mu_F^2}{m_H^2}\right)
\nonumber \\ 
\eeqa
\beqa
C_1^{(3)}&=&-160 C_F^3 \zeta_3 \ln\left(\frac{\mu_F^2}{m_H^2}\right)
+4 C_F^3 \ln^3\left(\frac{\mu_F^2}{m_H^2}\right)
\left[\ln\left(\frac{t_2 \, u_2}{m_H^4}\right)-\frac{3}{2}\right]
\nonumber \\ &&
{}+C_F^2 \beta_0 \ln^2\left(\frac{\mu_F^2}{m_H^2}\right)
\left[2\ln\left(\frac{\mu_R^2}{m_H^2}\right)
-\ln\left(\frac{\mu_F^2}{m_H^2}\right)\right]
-40 C_F^3 \zeta_2 \ln^2\left(\frac{\mu_F^2}{m_H^2}\right)
\nonumber \\ &&
{}-24 C_F^3 \zeta_2 \ln^2\left(\frac{\mu_F^2}{m_H^2}\right)
\left[4\ln\left(\frac{t_2 \, u_2}{m_H^4}\right)
+\frac{10}{3}\ln\left(\frac{m_H^2}{s}\right)-1\right] \, .
\eeqa
In the above expressions, $\beta_0=(11 C_A-2 n_f)/3$.
Once again, the logarithmic terms involving the variables $t_2$ and $u_2$ 
in the above expression arise from the details of the 1PI resummation. 

\mysection{Total cross sections for $H^- W^+$ production}

\begin{figure}
\begin{center}
\includegraphics[width=11cm]{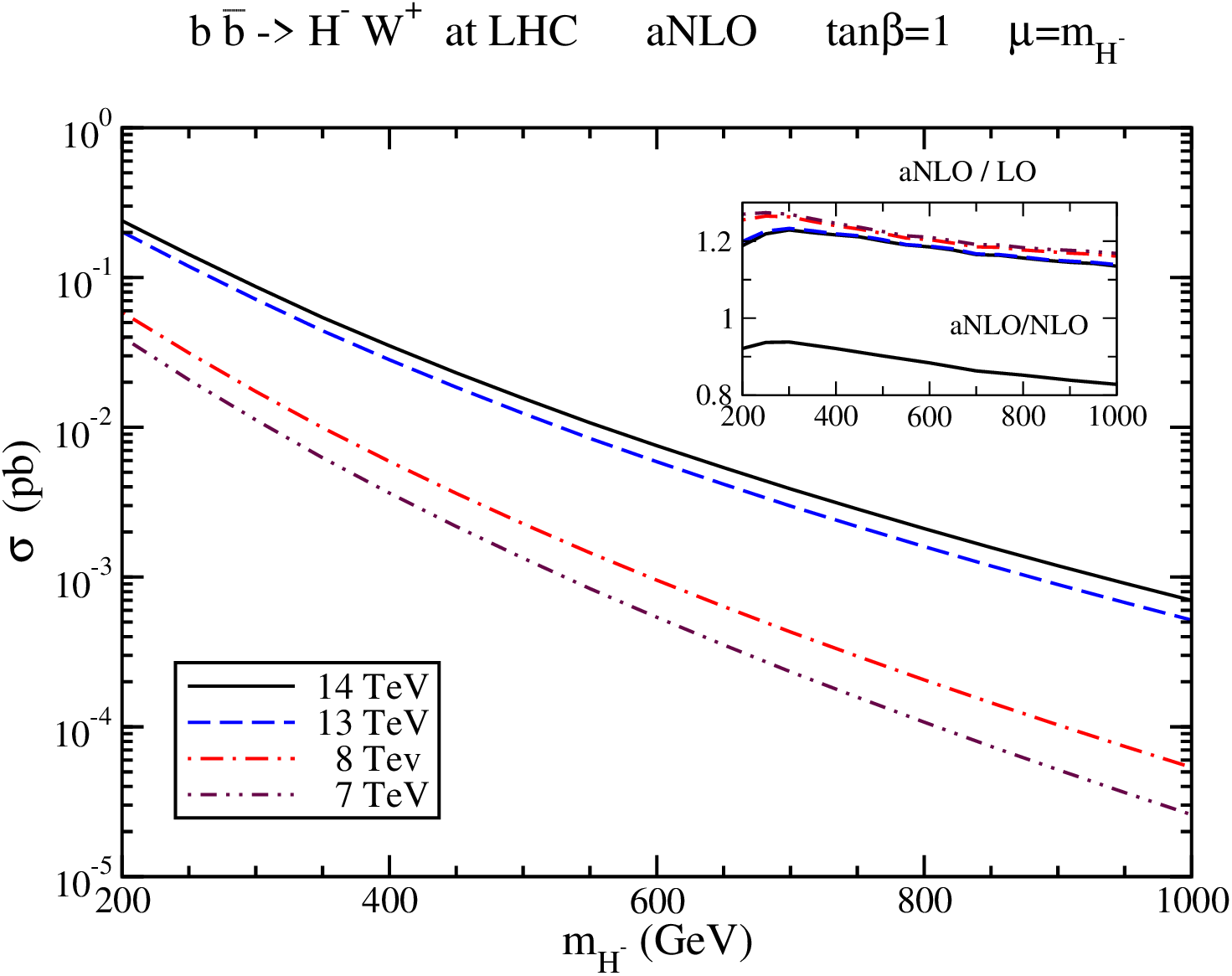}
\caption{The aNLO cross sections for 
$b{\bar b} \rightarrow H^- W^+$ at the LHC with 
$\sqrt{S}=7$, 8, 13, and 14 TeV.}
\label{bbHWaNLO}
\end{center}
\end{figure}

\begin{figure}
\begin{center}
\includegraphics[width=11cm]{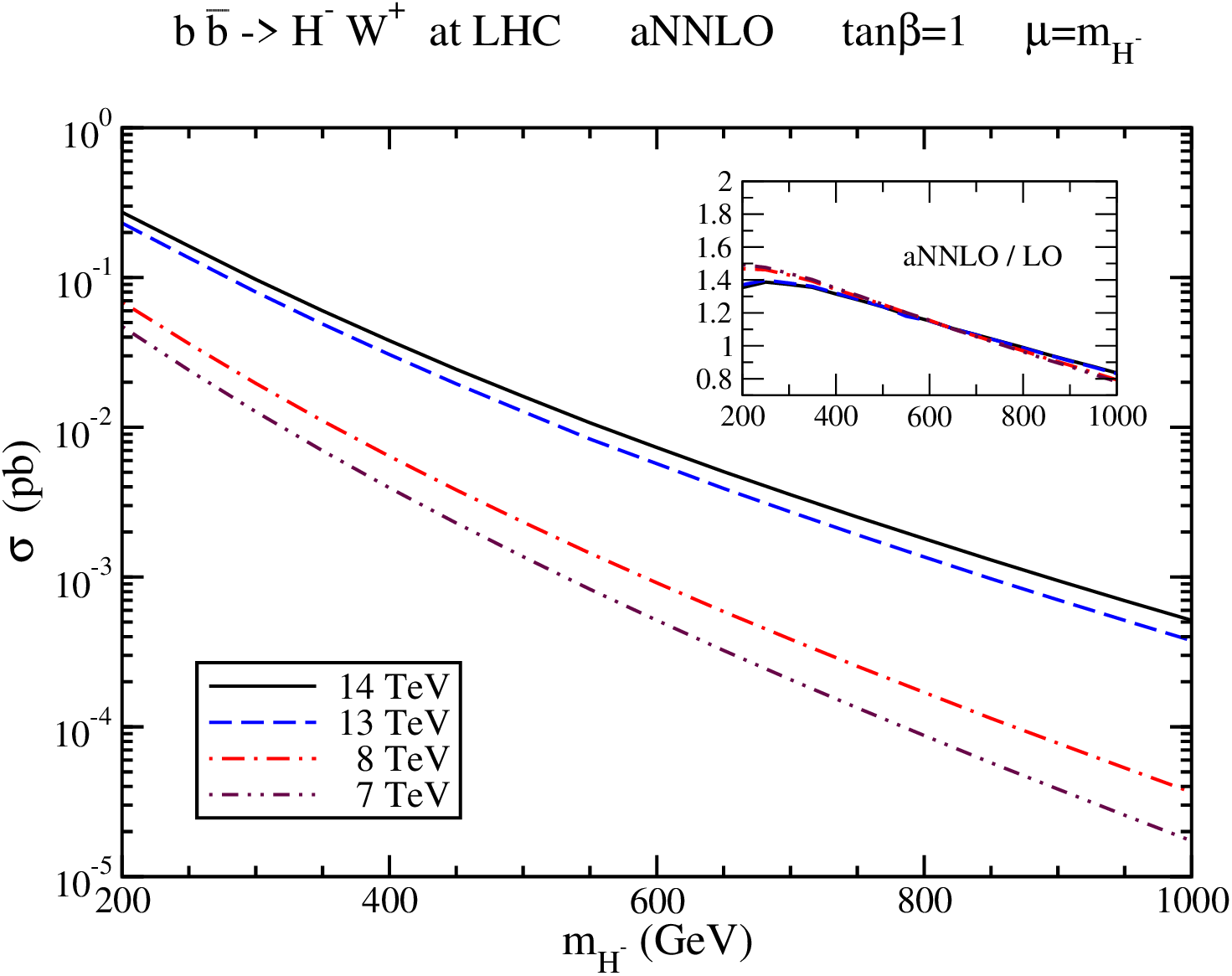}
\caption{The aNNLO cross sections for 
$b{\bar b} \rightarrow H^- W^+$ at the LHC with 
$\sqrt{S}=7$, 8, 13, and 14 TeV.}
\label{bbHWaNNLO}
\end{center}
\end{figure}

We consider proton-proton collisions with momenta 
$p(p_A)+p(p_B) \rightarrow H^-(p_3)+W^+(p_4)$. 
In analogy to the partonic variables defined in Section 2, we define the 
hadronic kinematical variables  
$S=(p_A+p_B)^2$, $T=(p_A-p_3)^2$, $T_1=T-m_H^2$, $T_2=T-m_W^2$, 
$U=(p_B-p_3)^2$, and $U_1=U-m_H^2$. The hadronic variables are related to the 
partonic variables via $p_1=x_1 p_A$ and $p_2=x_2 p_B$, where $x_1$ and $x_2$ 
are the fractions of the momentum carried by the partons in protons 
$A$ and $B$, respectively.

The hadronic total cross section can be written as 
\beqa
\sigma^{H^- W^+}&=& \int_{T^{min}}^{T^{max}} dT \int_{U^{min}}^{U^{max}} dU
\int_{x_2^{min}}^1 dx_2 \int_0^{s_4^{max}} ds_4 \, 
\frac{x_1 \, x_2}{x_2 S+T_1} \,
\phi(x_1) \, \phi(x_2) \, 
\frac{d^2{\hat\sigma}}{dt \, du}
\nonumber \\ &&
\label{totalcs}
\eeqa
where the $\phi$ denote the pdf; $x_1=(s_4-m_H^2+m_W^2-x_2U_1)/(x_2 S+T_1)$; 
$T^{^{max}_{min}}=-(1/2)(S-m_H^2-m_W^2) \pm (1/2) [(S-m_H^2-m_W^2)^2-4m_H^2m_W^2]^{1/2}$; 
$U^{max}=m_H^2+S m_H^2/T_1$ and $U^{min}=-S-T_1+m_W^2$;
$x_2^{min}=-T_2/(S+U_1)$; and $s_4^{max}=x_2(S+U_1)+T_2$.

Specifically, using the properties of plus distributions, Eq. (\ref{plus}), 
and the compact form of Eq. (\ref{NNLOcompact}), 
the aNNLO corrections to the total cross section, Eq. (\ref{totalcs}), 
can be written as 
\beqa
\sigma^{(2)}_{H^- W^+}&=& \frac{\alpha_s^2}{\pi^2} 
\int_{T^{min}}^{T^{max}} dT \int_{U^{min}}^{U^{max}} dU
\int_{x_2^{min}}^1 dx_2 \, \phi(x_2) \frac{x_2}{x_2 S+T_1}
\nonumber \\ && \hspace{-20mm}
\times \left\{-\int_0^{s_4^{max}} ds_4 \frac{1}{m_H^2} 
\ln^3\left(\frac{s_4}{m_H^2}\right) F_{LO} \, C_3^{(2)} \, x_1 \, \phi(x_1)
\right.
\nonumber \\ && \hspace{-15mm}
+\sum_{k=0}^3 \left[\int_0^{s_4^{max}} ds_4 
\frac{1}{s_4} \ln^k\left(\frac{s_4}{m_H^2}\right) 
\left(F_{\rm LO} \, C_k^{(2)} \, x_1 \, \phi(x_1)
-F_{\rm LO}^{\rm el} \, C_k^{(2) \rm el} \, x_1^{\rm el} \, \phi\left(x_1^{\rm el}
\right)\right) \right.
\nonumber \\ && \hspace{-15mm} \left. \left. 
{}+\frac{1}{k+1} \ln^{k+1}\left(\frac{s_4^{max}}{m_H^2}\right) 
F_{\rm LO}^{\rm el} \, C_k^{(2) \rm el} \, x_1^{\rm el} \, \phi\left(x_1^{\rm el}\right)  \right] \right\} 
\eeqa
where $x_1^{\rm el}$, $F_{\rm LO}^{\rm el}$, and $C_k^{(2)\, \rm el}$ denote the 
elastic variables, i.e. these quantities with $s_4=0$.
Analogous results can be written for the aNLO and aN$^3$LO corrections.

We now present results for the total $H^- W^+$ 
cross section at LHC energies using MMHT2014 NNLO pdf \cite{MMHT2014}. 
For convenience we set $\tan\beta=1$ but it is easy to rescale the results 
for any value of $\tan\beta$.

In Fig. \ref{bbHWaNLO} we plot the aNLO cross sections
for $b{\bar b} \rightarrow H^- W^+$ in proton-proton collisions at the LHC
versus charged Higgs mass for energies of 7, 8, 13, and 14 TeV. 
The cross sections vary greatly with charged Higgs mass, falling by three 
orders of magnitude over the mass range at each energy.
We also observe an order of magnitude or so increase in the cross section at 13 and 14 TeV relative to 7 and 8 TeV.

The inset plot of Fig. \ref{bbHWaNLO} shows the 
$K$-factors, i.e. the ratios of cross sections at various orders. 
The four lines at the top of the inset plot show the aNLO/LO ratios 
for the four LHC energies. The corrections are clearly very significant 
for all LHC energies. We also note that the $K$-factors at different energies 
are rather similar, and are slightly higher for smaller energies. 

It is also important to determine how much of the full NLO corrections 
\cite{HZ} are accounted for by the soft and collinear contributions. 
The lower line in the inset plot of Fig. \ref{bbHWaNLO} shows the aNLO/NLO 
ratio at 14 TeV energy. We see that the ratio is close to 1 for smaller 
charged-Higgs masses and it remains above 0.9 up to a mass of 500 GeV, 
indicating that the soft and collinear gluon corrections are dominant and 
provide numerically the majority of the NLO corrections. The ratio remains 
well above 0.8 through 1000 GeV, showing that the collinear and soft-gluon 
corrections are still large and significant.

In Fig. \ref{bbHWaNNLO} we plot the aNNLO cross sections
for $b{\bar b} \rightarrow H^- W^+$ versus charged Higgs mass 
for LHC energies of 7, 8, 13, and 14 TeV. 
Again, we observe a large increase in the cross section at 13 and 14 TeV 
relative to 7 and 8 TeV, and a large dependence of the cross section on the 
mass of the charged Higgs between 200 and 1000 GeV at each energy. 
The inset plot shows the aNNLO/LO $K$-factors.

We note that the leading collinear terms by themselves make a significant contribution to the total collinear plus soft corrections. For example, for 200 GeV charged Higgs mass at 13 TeV energy, they amount to 20\% of the aNNLO corrections.

Theoretical uncertainties arise from scale variation as well as from pdf uncertainties. Scale variation by a factor of 2 around the central scale $\mu=m_H$ produces a moderate uncertainty, $\pm 15$\% at 13 TeV LHC energy for a 500 GeV charged Higgs, with similar numbers at other energies. The uncertainties from the pdf are smaller, $\pm 5$\% at 13 TeV for a 500 GeV charged Higgs.

We find that results using other pdf sets are very similar. If one uses 
the CT14 NNLO pdf \cite{CT14} the results are essentially the same.

We note that the aN$^3$LO corrections are incomplete and their numerical contribution typically small relative to the aNLO and aNNLO corrections. For example, for 300 GeV charged-Higgs mass at 13 TeV energy, the aNLO corrections contribute a 23\% enhancement, the aNNLO corrections an additional 14\% enhancement, and the aN$^3$LO corrections a further 2\% enhancement. The fact that the aN$^3$LO corrections are much smaller than the corrections at previous orders is an indication of perturbative convergence, and is also in line with related results for Higgs production and top-quark production (see e.g. \cite{NKttbar}). Since the uncertainty due to uknown terms at aN$^3$LO can be of the order of the size of these corrections, we do not study them further. We also note that there are no pdf available at N$^3$LO for such calculations, and the effect of such pdf may also be nonnegligible.

\mysection{Charged Higgs $p_T$ and rapidity distributions}

\begin{figure}
\begin{center}
\includegraphics[width=11cm]{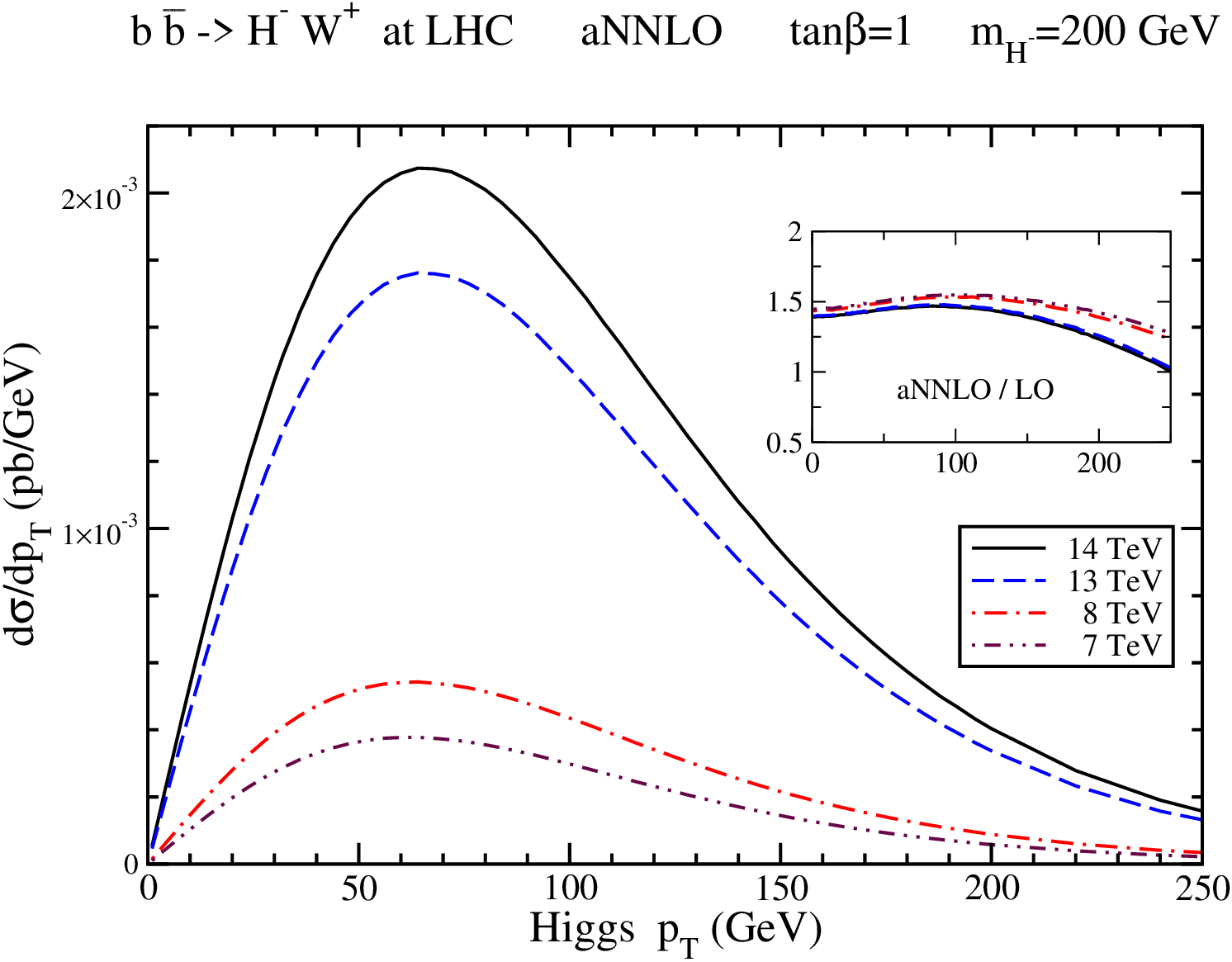}
\caption{The aNNLO charged Higgs $p_T$ distributions for 
$b{\bar b} \rightarrow H^- W^+$ at the LHC with $\sqrt{S}=7$, 8, 13, 
and 14 TeV,  and $m_H=200$ GeV.}
\label{ptbbHW200aNNLO}
\end{center}
\end{figure}

\begin{figure}
\begin{center}
\includegraphics[width=11cm]{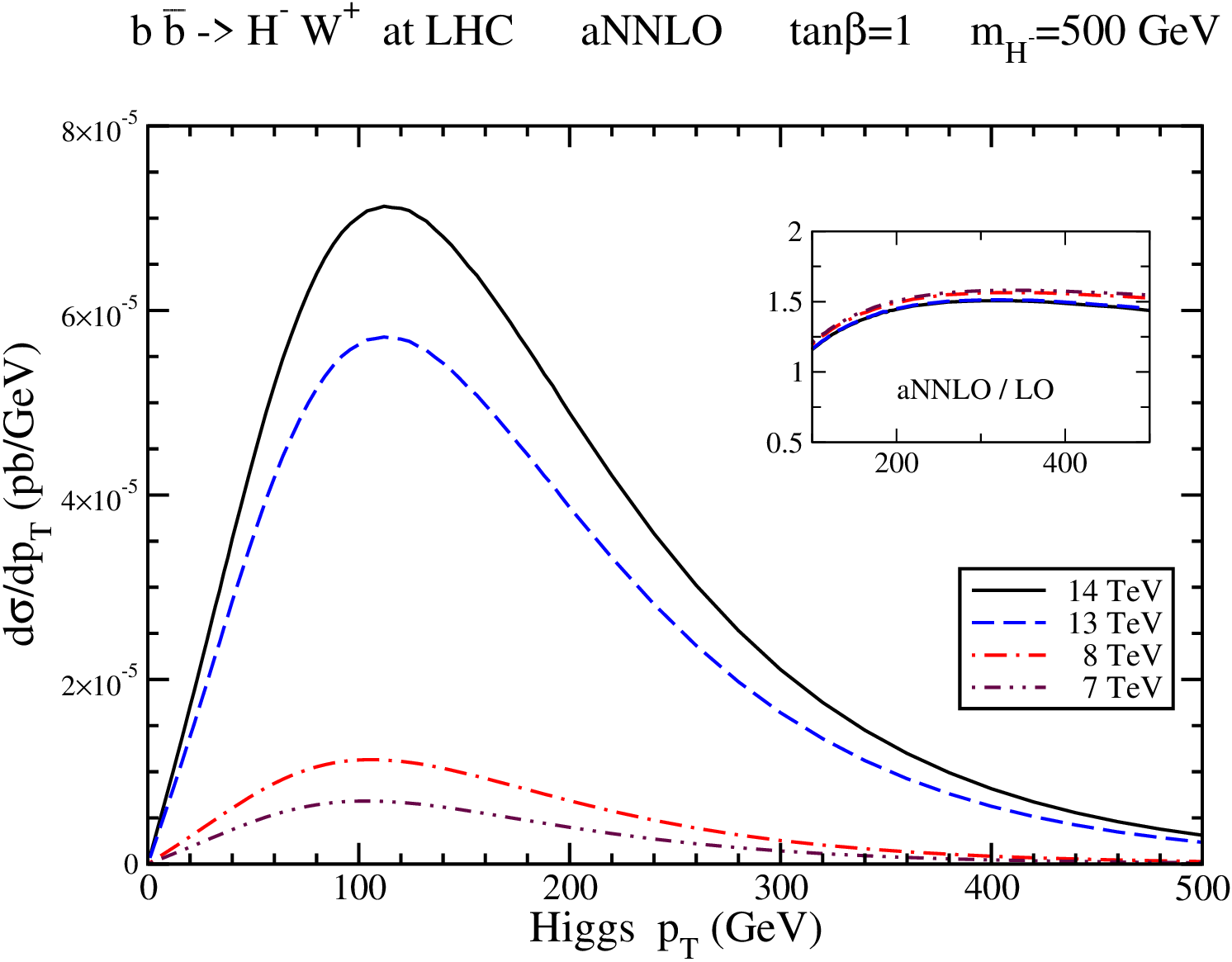}
\caption{The aNNLO charged Higgs $p_T$ distributions for 
$b{\bar b} \rightarrow H^- W^+$ at the LHC with $\sqrt{S}=7$, 8, 13, 
and 14 TeV,  and $m_H=500$ GeV.}
\label{ptbbHW500aNNLO}
\end{center}
\end{figure}

We continue with the charged Higgs $p_T$ and rapidity distributions.
The charged Higgs $p_T$ distribution is given by 
\beqa
\frac{d\sigma}{dp_T}&=& 2 \, p_T \int_{Y^{min}}^{Y^{max}} dY
\int_{x_2^{min}}^1 dx_2 \int_0^{s_4^{max}} ds_4 \, 
\frac{x_1 \, x_2 \, S}{x_2 S+T_1} \, \phi(x_1) \, \phi(x_2) \, 
\frac{d^2{\hat\sigma}}{dt \, du}
\label{pTdist}
\eeqa
where $T_1=-\sqrt{S} \, (m_H^2+p_T^2)^{1/2} \, e^{-Y}$,  
$U_1=-\sqrt{S} \, (m_H^2+p_T^2)^{1/2} \, e^{Y}$,  
$Y^{^{max}_{min}}=\pm (1/2) \ln[(1+\beta_T)/(1-\beta_T)]$ with 
$\beta_T=[1-4(m_H^2+p_T^2)S/(S+m_H^2-m_W^2)^2]^{1/2}$, and the other quantities 
are defined in Section 3. We note that the total cross section can also be 
calculated by integrating the $p_T$ distribution, $d\sigma/dp_T$, 
over $p_T$ from 0 to $p_T^{max}=[(S-m_H^2-m_W^2)^2-4m_H^2m_W^2]^{1/2}/(2\sqrt{S})$, 
and we have checked for consistency that we get the same numerical results 
as in Section 3.

In Fig. \ref{ptbbHW200aNNLO} we plot the aNNLO $p_T$ distributions, $d\sigma/dp_T$, of the charged Higgs boson with mass 200 GeV for LHC energies of 7, 8, 13, and 14 TeV. The inset plot shows the aNNLO/LO $K$-factors. The corrections are large, around 50\%, for much of the $p_T$ range shown. The distributions peak at a $p_T$ value of around 65 GeV for this mass choice.

In Fig. \ref{ptbbHW500aNNLO} we plot the corresponding aNNLO $p_T$ distributions of the charged Higgs boson with mass 500 GeV. The inset plot shows the aNNLO/LO $K$-factors and, again, the corrections are large. The distributions now peak at a higher $p_T$ value of around 110 GeV.

\begin{figure}
\begin{center}
\includegraphics[width=11cm]{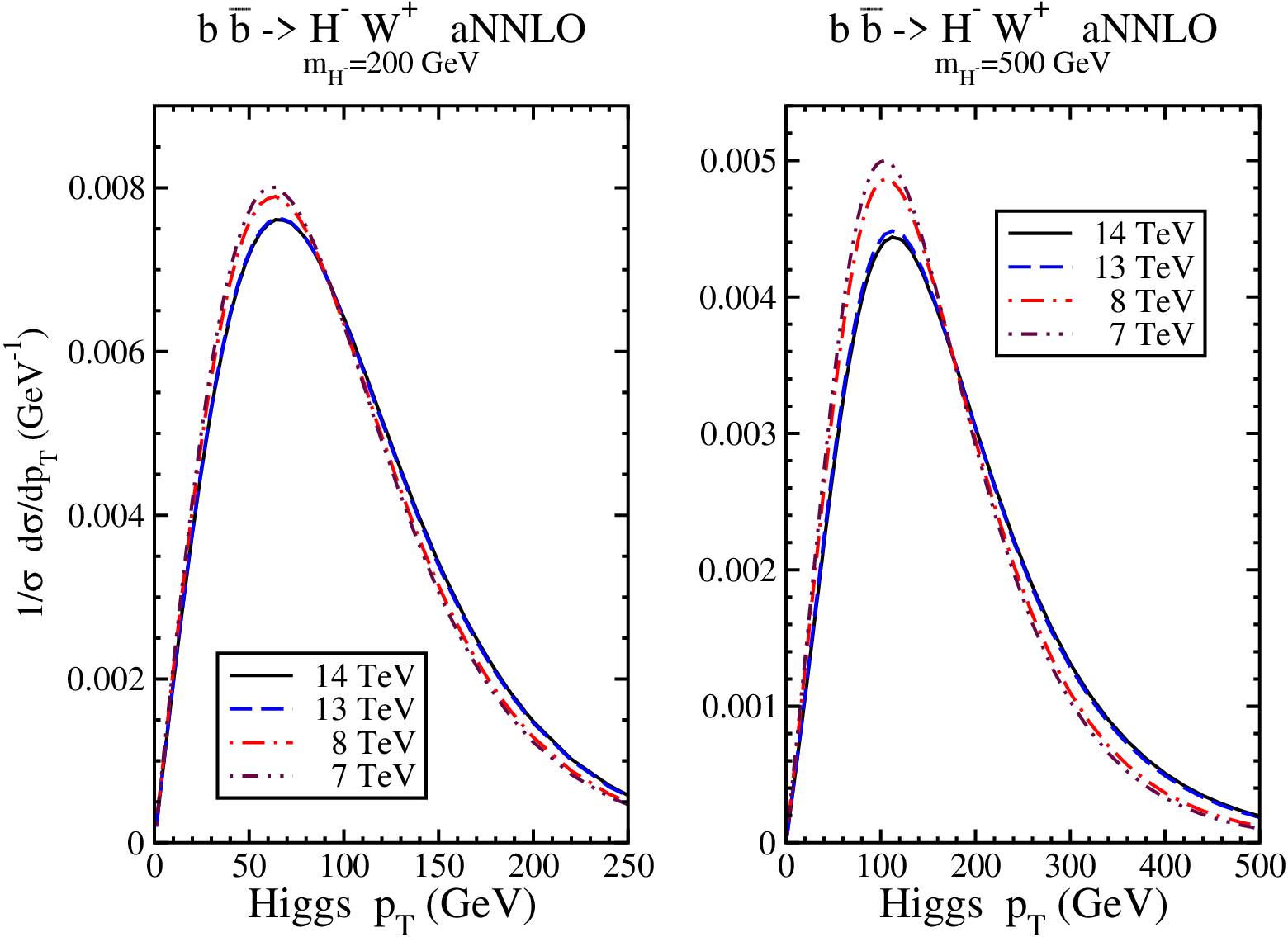}
\caption{The aNNLO charged Higgs normalized $p_T$ distributions for 
$b{\bar b} \rightarrow H^- W^+$ at the LHC with $\sqrt{S}=7$, 8, 13, 
and 14 TeV,  and $m_H=200$ GeV (left) and 500 GeV (right).}
\label{ptnormbbHWaNNLO}
\end{center}
\end{figure}

It is useful to also study normalized distributions since normalization removes the dependence on $\tan\beta$ and it minimizes the dependence on the choice of pdf. Such normalized distributions are also often favored in experimental studies and comparisons with theory.

In Fig. \ref{ptnormbbHWaNNLO} we plot the aNNLO normalized $p_T$ distributions, $(1/\sigma) d\sigma/dp_T$, of the charged Higgs boson with mass 200 GeV (left plot) and 500 GeV (right plot) for LHC energies of 7, 8, 13, and 14 TeV. The shape of the normalized $p_T$ distributions depends on the energy, as expected, with higher peaks at lower energies. We also observe that the peaks are lower for a 500 GeV mass than for 200 GeV. 

The charged-Higgs rapidity, $Y$, distribution is given by 
\beqa
\frac{d\sigma}{dY}&=& \int_0^{p_T^{max}} 2 \, p_T \, dp_T
\int_{x_2^{min}}^1 dx_2 \int_0^{s_4^{max}} ds_4 \, 
\frac{x_1 \, x_2 \, S}{x_2 S+T_1} \, \phi(x_1) \, \phi(x_2) \, 
\frac{d^2{\hat\sigma}}{dt \, du}
\label{Ydist}
\eeqa
where $p_T^{max}=((S+m_H^2-m_W^2)^2/(4S\cosh^2Y)-m_H^2)^{1/2}$ and the rest of the quantities are defined as before. We again note that the total cross section can also be obtained by integrating the rapidity distribution, $d\sigma/dY$, over rapidity with limits $Y^{^{max}_{min}}=\pm (1/2) \ln[(1+\beta)/(1-\beta)]$ where $\beta=(1-4m_H^2/S)^{1/2}$, and again we have checked for consistency that we get the same numerical results as in Section 3.

\begin{figure}
\begin{center}
\includegraphics[width=11cm]{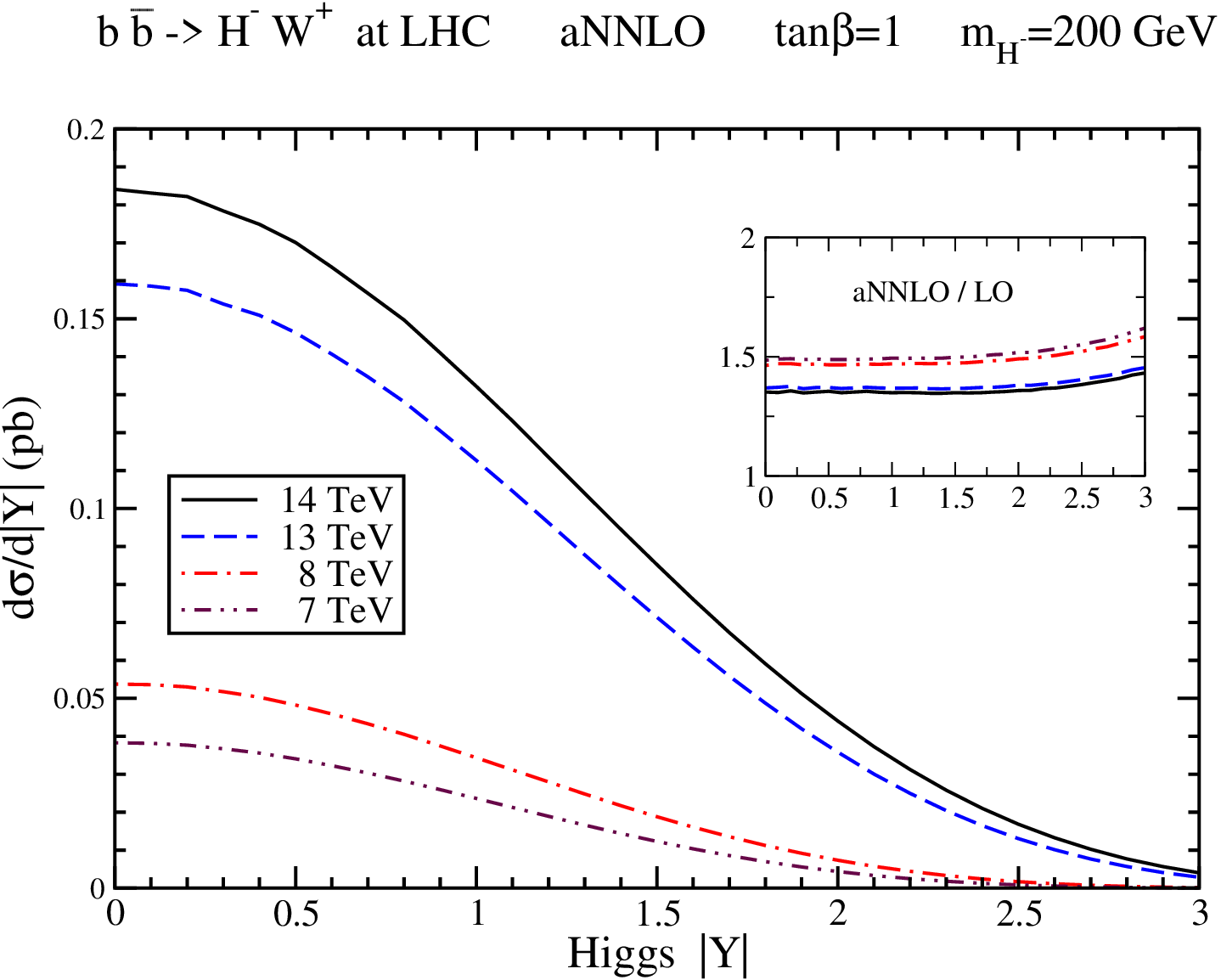}
\caption{The aNNLO charged Higgs rapidity distributions for 
$b{\bar b} \rightarrow H^- W^+$  at the LHC with 
$\sqrt{S}=7$, 8, 13, and 14 TeV, and $m_H=200$ GeV.}
\label{yabsbbHW200aNNLO}
\end{center}
\end{figure}

In Fig. \ref{yabsbbHW200aNNLO} we plot the aNNLO rapidity distributions, 
$d\sigma/d|Y|$, of the charged Higgs boson with mass 200 GeV for LHC energies 
of 7, 8, 13, and 14 TeV. The inset plot shows the aNNLO/LO $K$-factors.
The corrections are quite large, especially at lower LHC energies, and 
they grow at larger values of charged Higgs rapidity.

\begin{figure}
\begin{center}
\includegraphics[width=11cm]{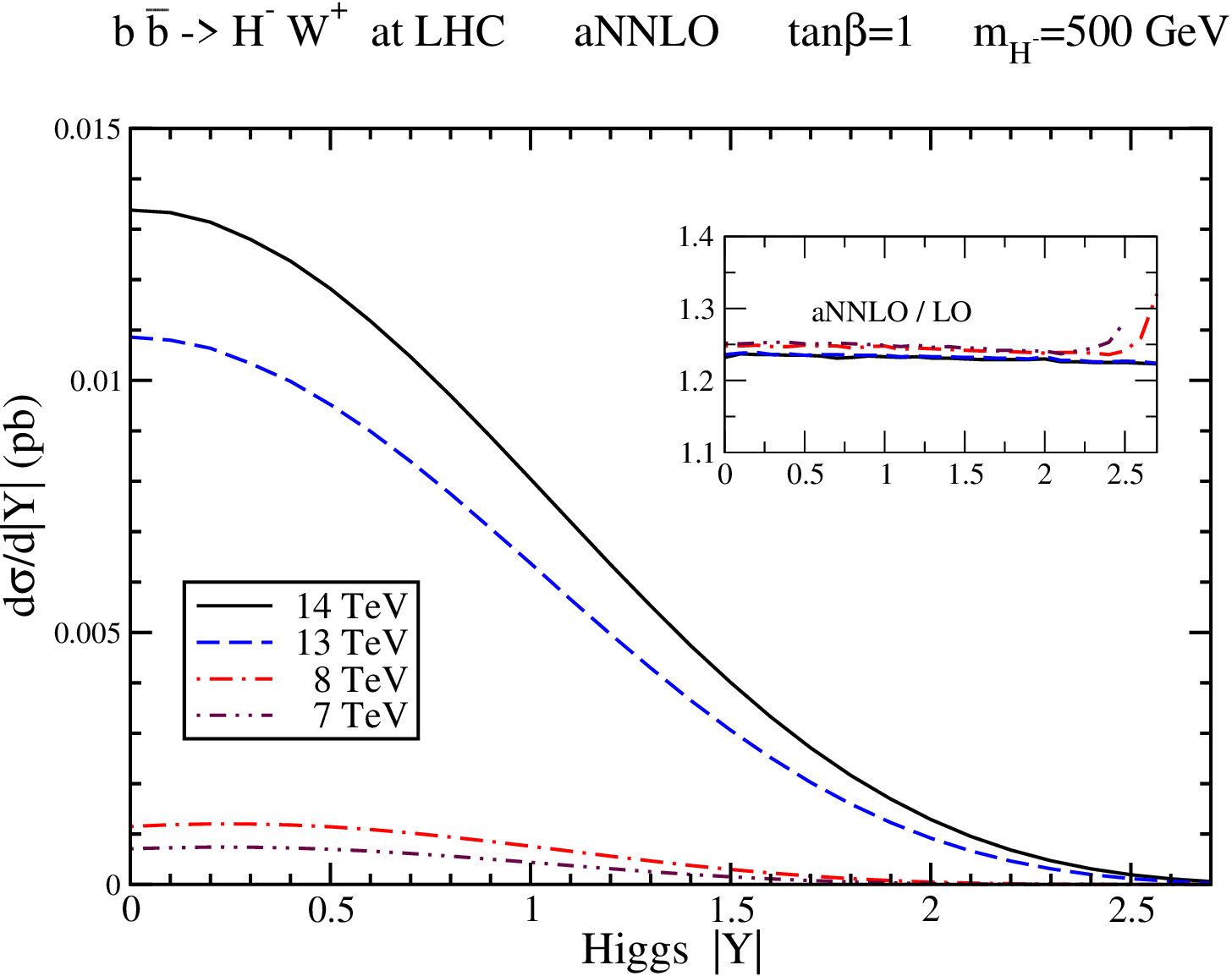}
\caption{The aNNLO charged Higgs rapidity distributions for 
$b{\bar b} \rightarrow H^- W^+$  at the LHC with 
$\sqrt{S}=7$, 8, 13, and 14 TeV, and $m_H=500$ GeV.}
\label{yabsbbHW500aNNLO}
\end{center}
\end{figure}

In Fig. \ref{yabsbbHW500aNNLO} we plot the corresponding aNNLO rapidity distributions of the charged Higgs boson with mass 500 GeV. The aNNLO/LO $K$-factors are again shown in the inset plot. We observe that the 7 and 8 TeV $K$-factors increase rapidly at larger values of rapidity.

\begin{figure}
\begin{center}
\includegraphics[width=11cm]{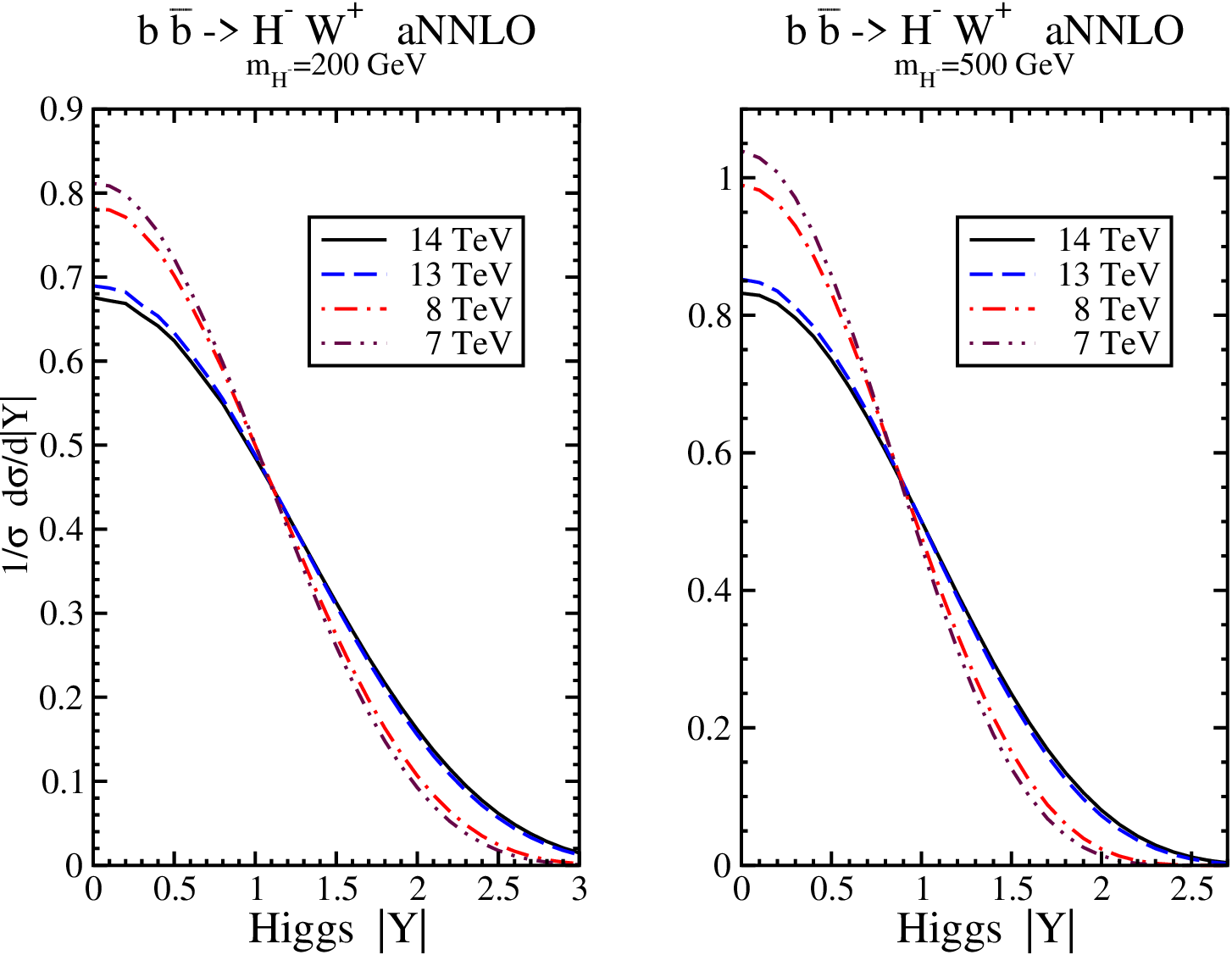}
\caption{The aNNLO charged Higgs normalized rapidity distributions for 
$b{\bar b} \rightarrow H^- W^+$  at the LHC with 
$\sqrt{S}=7$, 8, 13, and 14 TeV, and $m_H=200$ GeV (left)  
and 500 GeV (right).}
\label{ynormabsbbHWaNNLO}
\end{center}
\end{figure}

Finally, in Fig. \ref{ynormabsbbHWaNNLO} we plot the aNNLO normalized rapidity distributions, $(1/\sigma) d\sigma/d|Y|$, of the charged Higgs boson with mass 200 GeV (left plot) and 500 GeV (right plot) for LHC energies 
of 7, 8, 13, and 14 TeV. For a given charged Higgs mass the normalized rapidity
distributions at lower energies have higher peaks at central rapidity with corresponding smaller values at large $|Y|$, as expected. The fall of the distributions with increasing $|Y|$ is sharper for $m=500$ GeV than for 200 GeV at all LHC energies.

\mysection{Conclusions}

The cross sections for the associated production of a charged Higgs boson with 
a $W$ boson, via $b{\bar b} \rightarrow H^- W^+$, receive sizable contributions 
from collinear and soft gluon corrections. 
These radiative contributions have been resummed, and approximate  
double-differential cross sections have been derived at NLO, NNLO, and N$^3$LO. 
Numerical predictions have been provided for the total cross section for 
$H^- W^+$ production at LHC energies as well as for the $p_T$ and rapidity 
distributions of the charged Higgs boson. 
The higher-order corrections are significant and they enhance the 
total cross section and differential distributions for $H^- W^+$ production 
at the LHC.

\mysection*{Acknowledgements}
This material is based upon work supported by the National Science Foundation under Grant No. PHY 1519606.

\end{document}